\documentclass{elsart5p}
\usepackage[tbtags]{amsmath}
\usepackage{graphicx}

\def\D{d}
\def\LRVec#1{\mathsf{#1}}
\advance\voffset-45pt
\advance\hoffset8.075pt

\begin{document}

\journal{Phys.\ Lett.\ A}
\date{\textup{SAND2007-1087J}}

\begin{frontmatter}

\title{Scaling of Huygens-front speedup in weakly random media}

\author{Jackson R. Mayo\corauthref{cor}},
\corauth[cor]{Corresponding author.}
\ead{jmayo@sandia.gov}
\author{Alan R. Kerstein}
\ead{arkerst@sandia.gov}

\address{Combustion Research Facility, Sandia National Laboratories, Livermore,
CA 94551, USA}

\begin{abstract}
Front propagation described by Huygens' principle is a fundamental mechanism of
spatial spreading of a property or an effect, occurring in optics, acoustics,
ecology and combustion. If the local front speed varies randomly due to
inhomogeneity or motion of the medium (as in turbulent premixed combustion),
then the front wrinkles and its overall passage rate (turbulent burning
velocity) increases. The calculation of this speedup is subtle because it
involves the minimum-time propagation trajectory. Here we show mathematically
that for a medium with weak isotropic random fluctuations, under mild
conditions on its spatial structure, the speedup scales with the 4/3 power of
the fluctuation amplitude. This result, which verifies a previous conjecture
while clarifying its scope, is obtained by reducing the propagation problem to
the inviscid Burgers equation with white-in-time forcing. Consequently,
field-theoretic analyses of the Burgers equation have significant implications
for fronts in random media, even beyond the weak-fluctuation limit.
\end{abstract}


\end{frontmatter}

\section{\label{Intro}Introduction}

Phenomena from combustion \cite{W85} to seismic waves \cite{SP99} to population
spreading \cite{M02} can be modeled by Huygens' principle of front propagation,
first stated as a law of geometrical optics. In each application, the boundary
of the affected region (idealized as a sharp front) advances normal to itself
at a locally specified speed. In a uniform medium, where this speed is
constant, an initially wrinkled front flattens out over time; but in a
spatially varying medium, a competition occurs as wrinkling is continually
reintroduced \cite{KA92}. A central problem for the latter case is to determine
the overall statistically steady propagation rate of the wrinkled front, which
exceeds the average local speed because the front is defined by the fastest
paths through the medium (first passage). Our main result, establishing under
general conditions the proportionality of the speedup to the 4/3 power of the
amplitude of weak fluctuations, agrees with previous heuristic analysis
\cite{KA92} and numerical simulations \cite{KA92,RMS93,KA94,MSS97}, and
eliminates ambiguities \cite{AB03} concerning the physical relevance of the
scaling. In the important case of premixed fluid combustion, the result
describes the weak-advection scaling of the turbulent burning velocity
\cite{F95}.

Our analysis relates the front dynamics to the evolution of a pressure-free
fluid obeying the white-noise-driven Burgers equation, itself a widely studied
model of turbulence \cite{P95}. The results reported here enable adaptation of
existing treatments of Burgers turbulence \cite{BMP95,GK98} to estimate the
prefactor of the speedup scaling and its dependence on medium structure, with
implications even for the opposite, practical limit of strongly advected
flames.

The propagation of a Huygens front in a general nonuniform medium is governed
by the local advecting velocity of the medium as a function of time and space,
$\mathbf{u}(t,\mathbf{x})$, and the local speed of propagation relative to the
medium, $v(t,\mathbf{x})$. Huygens' principle can then be stated as follows
(using three-dimensional language for definiteness): If at time~$t_0$ a
point~$\mathbf{x}_0$ lies in the affected region (including its boundary, the
front), then at a slightly later time $t_1 = t_0 + \D t$ all points in a ball
of radius $v(t_0,\mathbf{x}_0)\, \D t$ about the point $\mathbf{x}_0 +
\mathbf{u}(t_0,\mathbf{x}_0)\, \D t$ are affected. The boundary of the affected
region at~$t_1$ (the new front) thus consists of certain points on the surfaces
of balls originating from the initial front. If we consider all points on these
spherical surfaces, then we automatically include those on the new front and
more. Hence the front at any later time will be found among the affected points
reached by arbitrary trajectories $\mathbf{x}(t)$ that start on the initial
front and always move at the local speed $v$ (in any direction) relative to the
medium, so that they obey
\begin{equation}
\label{huygens}
\left|\frac{\D\mathbf{x}}{\D t} - \mathbf{u}(t,\mathbf{x})\right| =
v(t,\mathbf{x}).
\end{equation}

Two simplifications of this general framework are physically important. First,
for geometrical optics in a static or ``quenched'' medium with nonuniform
refractive index, or for combustion of a solid propellant with nonuniform
burning rate, we have $v = v(\mathbf{x})$ (time-independent fluctuations) and
$\mathbf{u} \equiv \mathbf{0}$ (no advection). Note that our nonrelativistic
equations do not correctly describe advection of light anyway, but are adequate
for an optical medium ``at rest'' (where fluctuations of~$v$ dominate those
of~$\mathbf{u}$) and for advection of sound in geometrical acoustics. Second,
for idealized combustion of a premixed turbulent fluid in the limit of a very
thin flame front \cite{W85}, $v$~is a constant (the laminar flame speed) and
$\mathbf{u}(t,\mathbf{x})$ is the turbulent flow (assumed to be unaffected by
the flame). Following the derivation of our key results in Sections
\ref{Fronts} and \ref{Weak}, further implications for combustion are discussed
in Section~\ref{Concl}.

\section{\label{Fronts}Fronts and particles}

In the general case, Eq.\ (\ref{huygens}) defines a large family of ``virtual''
trajectories $\mathbf{x}(t)$ of which only a subset actually form the front at
a given time. The criterion for the relevant trajectories is simplest if
$|\mathbf{u}| < v$ everywhere, as we now assume (in Section~\ref{Concl} we
discuss relaxing this assumption). This inequality, which is trivially
satisfied for a quenched medium, ensures that advection can never sweep the
front backward and thus that each point $\mathbf{y}$ is crossed by the front
only once. The time of this crossing, which we call $T_0(\mathbf{y})$, is
simply the time when the first virtual trajectory reaches $\mathbf{y}$, hence
defining a first-passage problem.

To reduce the number of extraneous trajectories, we can exploit the
first-passage criterion (minimization of travel time) and obtain constraints on
relevant trajectories. Let us parametrize the solutions of Eq.\ (\ref{huygens})
by
\begin{equation}
\label{dxdt}
\frac{\D\mathbf{x}}{\D t} = \mathbf{u} + v \mathbf{n},
\end{equation}
with $\mathbf{n}(t)$ a unit vector. Then (see Appendix~\ref{Law}) a necessary
condition for a first-passage trajectory is
\begin{equation}
\label{dndt}
\frac{\D\mathbf{n}}{\D t} = -P_{\mathbf{n}}(\LRVec A \cdot \mathbf{n} +
\boldsymbol{\nabla}v),
\end{equation}
where $P_{\mathbf{n}}$ denotes the projection orthogonal to~$\mathbf{n}$ and
$\LRVec A$~is the velocity gradient tensor $A_{ij} = \nabla_i u_j$; a further
necessary condition is that the trajectory starts out with $\mathbf{n}$ normal
to the initial front, from which it follows that $\mathbf{n}$ remains normal to
the evolving front. In a quenched medium, where $\mathbf{u}$ and $\LRVec A$ are
zero, Eqs.\ (\ref{dxdt}) and (\ref{dndt}) reduce to the ray equations of
geometrical optics, and there it is well known that rays propagate normal to
fronts. But with advection, we see that a trajectory's tangent vector
$\D\mathbf{x}/\D t$ is no longer aligned with the front normal~$\mathbf{n}$. As
discussed in Section~\ref{Concl}, Eqs.\ (\ref{dxdt}) and (\ref{dndt}) govern
front-tracking trajectories even when $|\mathbf{u}| \ge v$, but we assume
$|\mathbf{u}| < v$ so that the front evolution is described by a simple
first-passage problem and a single-valued function $T_0(\mathbf{y})$.

For first-passage purposes, then, we consider a continuum of ``particles''
starting simultaneously from all points on the initial front (with initial
$\mathbf{n}$ given by the unit normal), and obeying Eqs.\ (\ref{dxdt}) and
(\ref{dndt}). If the Huygens front represents a wave phenomenon in the
geometrical-optics limit of very short wavelength, then we are describing
physical quasiparticles: photons for light or phonons for sound. Motivated by
this physical case, we call Eqs.\ (\ref{dxdt}) and (\ref{dndt}) the ``law of
motion'' for first-passage trajectories. In premixed combustion, where the
front represents a thin flame, the ``particle'' trajectories are mathematical
constructs known as ignition curves \cite{S85}.

\begin{figure}
\begin{center}
\includegraphics[width=88.5mm]{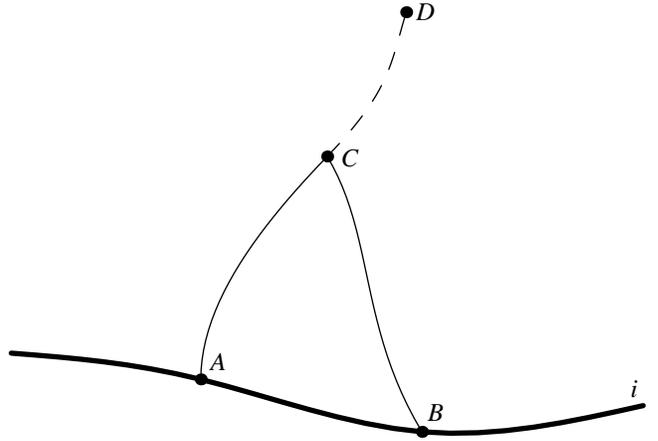}%
\end{center}
\caption{\label{collision}Irrelevance of particles after collision. Two
particles, starting from $A$ and $B$ on the initial front $i$, collide at $C$;
the continuation of trajectory $AC$ (dashed) reaches $D$. If it were the first
arrival, its travel time would be an absolute minimum over all virtual
trajectories from $i$ to $D$, including $BCD$. Having the same minimum time,
$BCD$ must also obey the law of motion, but this deterministic law does not
permit the time-reversed trajectory $DC$ to split at $C$. Hence neither $ACD$
nor $BCD$ represents first passage to $D$, and the particles can be discarded
upon colliding.}
\end{figure}

Even with the condition (\ref{dndt}), not all of these particles remain on the
front. The departure of particles from the front (into the interior of the
affected region) is associated with ``cusps'' at which the normal $\mathbf{n}$
is not unique. Such cusps develop during propagation in a random medium even if
the initial front is smooth \cite{W84,KAW88}. When two distinct particles reach
the same point at the same time (necessarily with different $\mathbf{n}$), both
colliding particles fall behind the front and can then be discarded, as shown
in Fig.~\ref{collision}. (In optics and acoustics, the corresponding photons
and phonons remain observable as second and later arrivals, but our scope is
limited to first passage.)

The collision rule implies a correspondence between our continuum of particles
and a model ``fluid'' without pressure or viscosity, consisting of fluid
elements that move independently in response to external forces but disappear
when they collide. Such a fluid is described by the inviscid Burgers equation
(with suitable forcing) and the collisions are known as shocks \cite{S85,B05}.
Because our particles fill a surface and not all of space, we must define the
corresponding Burgers fluid more precisely. The particles, by definition,
always represent the first arrival at their locations; thus if two of their
paths reach the same point~$\mathbf{x}$, the particles necessarily arrive at
the same time $T_0(\mathbf{x})$ and are then discarded. Let us take the initial
front to be planar and use coordinates $\mathbf{x} =
(x_\parallel,\mathbf{x}_\perp)$ such that this plane is $x_\parallel = 0$. Then
the spatial paths of the particles can be viewed as ``trajectories''
$\mathbf{x}_\perp(x_\parallel)$ in the ``time''~$x_\parallel$.

In this picture, a Burgers fluid exists in the lower-dimensional space
$\mathbf{x}_\perp$, and the physical time information is carried by the
function $T_0(x_\parallel,\mathbf{x}_\perp)$. Here we must introduce the
assumption that the medium fluctuations are weak, i.e., that $|\mathbf{u}|$ and
the changes in~$v$ are both vanishingly small compared to the average value
of~$v$. In this limit, because trajectories cannot deviate significantly from
the $x_\parallel$-direction without falling behind the front, the particles
necessarily collide before they are deflected enough to make the function
$\mathbf{x}_\perp(x_\parallel)$ ill defined.

\begin{figure}
\begin{center}
\includegraphics[width=88.5mm]{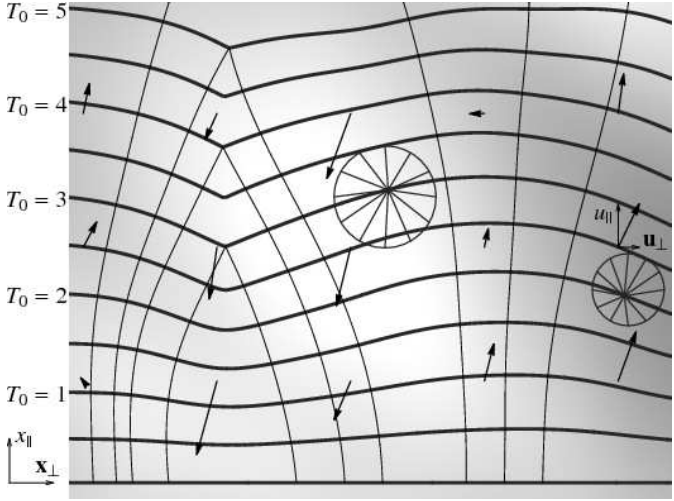}%
\end{center}
\caption{\label{eikonal}Complementary descriptions of first passage. A front
propagates upward amid possibly time-dependent variations in $\mathbf{u}$
(arrows) and $v$ (grayscale, dark for slow). We show $\mathbf{u}$ and $v$ at
the local time $T_0$. Snapshots of the front (thick curves) are contours of
$T_0$; particles (thin curves) track the front until they collide at cusps. Two
circles expanding from affected points illustrate Huygens' principle. The
fluctuations are weak enough to define a ``fluid'' using $x_\parallel$ as
``time'' for the particles. For very weak fluctuations, the front remains
nearly flat, and $\mathbf{u}_\perp$ becomes irrelevant as it merely shifts the
front without advancing it.}
\end{figure}

The law of motion for these trajectories $\mathbf{x}_\perp(x_\parallel)$ could
be derived from Eqs.\ (\ref{dxdt}) and (\ref{dndt}), but it is simpler to
obtain the Burgers equation in its conventional Eulerian form. The same
minimization principle that yields Eqs.\ (\ref{dxdt}) and (\ref{dndt}) also
implies that $T_0(\mathbf{x})$ satisfies a generalized ``eikonal'' equation
(see Appendix~\ref{Eikonal})
\begin{equation}
\label{eik}
\mathbf{u}\bigl(T_0(\mathbf{x}),\mathbf{x}\bigr) \cdot
\boldsymbol{\nabla}T_0(\mathbf{x}) + v\bigl(T_0(\mathbf{x}),\mathbf{x}\bigr)\,
|\boldsymbol{\nabla}T_0(\mathbf{x})| = 1.
\end{equation}
The relation between $T_0$ and the particle trajectories is illustrated in
Fig.~\ref{eikonal} for a two-dimensional medium. Because $T_0$ is constant over
a front, $\boldsymbol{\nabla}T_0$ lies along the front normal~$\mathbf{n}$ (in
the direction of propagation), and so the physical velocity of particles in
Eq.\ (\ref{dxdt}) can be written
\begin{equation}
\label{dxdt1}
\frac{\D\mathbf{x}}{\D t} = \mathbf{u} + v\,
\frac{\boldsymbol{\nabla}T_0}{|\boldsymbol{\nabla}T_0|}.
\end{equation}
We next show that in the limit of weak random fluctuations, Eq.\ (\ref{eik})
reduces to the forced inviscid Burgers equation for a fluid consisting of these
particles.

\section{\label{Weak}Weak-fluctuation limit}

In a reference frame where the average value of~$\mathbf{u}$ is zero, and in
units such that the average value of~$v$ is unity, let us parametrize the weak
fluctuations by
\begin{align}
\label{uU}
\mathbf{u}(t,\mathbf{x}) &= \epsilon\, \mathbf{U}(\epsilon
t,x_\parallel,\mathbf{x}_\perp),\\
\label{vV}
v(t,\mathbf{x}) &= 1 + \epsilon\, V(\epsilon t,x_\parallel,\mathbf{x}_\perp),
\end{align}
where $\mathbf{U}$ and $V$ are homogeneous isotropic random fields and
$\epsilon$ is taken asymptotically to zero. The time dependence of $\mathbf{u}$
and~$v$ is scaled by $\epsilon$ because the natural source of time dependence
is advection by $\mathbf{u}$ itself, which goes to zero with~$\epsilon$.
Heuristic scaling analysis \cite{KA92,KA94} yields the following conclusions in
the $\epsilon \to 0$ limit: The medium can be considered effectively frozen
($\mathbf{U}$ and $V$ time-independent); the advection component
$\mathbf{U}_\perp$ orthogonal to the overall propagation direction is
irrelevant; the front reaches a statistically steady state over a distance of
order $\epsilon^{-2/3}$; and the steady passage rate exceeds unity by an amount
of order $\epsilon^{4/3}$.

Guided by these expectations, we define rescaled quantities
\begin{align}
\label{xi}
\xi &= \epsilon^p x_\parallel,\\
\label{tau}
\tau(\xi,\mathbf{x}_\perp) &= \epsilon^{-p} [T_0(x_\parallel,\mathbf{x}_\perp)
- x_\parallel],
\end{align}
where we anticipate that $p = \tfrac{2}{3}$ produces a useful $\epsilon \to 0$
limit, but we also consider slightly different $p$~values to see why
$\tfrac{2}{3}$ is special. The point of the rescaling is that if $\nabla_\xi
\tau$ reaches a steady-state average value $-C$ after a finite-$\xi$ transient,
then $\nabla_\parallel T_0$ averages to $1 - C\epsilon^{2p}$ after a
characteristic distance $x_\parallel \sim \epsilon^{-p}$. A planar front
propagating at constant speed $v_*$ would have $\nabla_\parallel T_0 = 1/v_*$,
and so to leading order the passage rate in the random medium is
\begin{equation}
v_* = 1 + C\epsilon^{2p}.
\end{equation}
Our main result is a demonstration that for $p = \tfrac{2}{3}$, the rescaled
front is governed by a forced Burgers equation that reaches such a steady
state, thus implying $\epsilon^{4/3}$ dependence of the speedup.

Substituting Eqs.\ (\ref{uU})--(\ref{tau}) into the eikonal equation
(\ref{eik}), we find
\begin{equation}
\begin{split}
&\epsilon\, U_\parallel(\epsilon^{1-p} \xi + \epsilon^{1+p} \tau, \epsilon^{-p}
\xi, \mathbf{x}_\perp)\, (1 + \epsilon^{2p} \nabla_\xi \tau)\\
&\qquad + \epsilon\, \mathbf{U}_\perp(\epsilon^{1-p} \xi + \epsilon^{1+p} \tau,
\epsilon^{-p} \xi, \mathbf{x}_\perp) \cdot \epsilon^p \boldsymbol{\nabla}_\perp
\tau\\
&\qquad + [1 + \epsilon\, V(\epsilon^{1-p} \xi + \epsilon^{1+p} \tau,
\epsilon^{-p} \xi, \mathbf{x}_\perp)]\\
&\qquad\qquad \times \sqrt{(1 + \epsilon^{2p} \nabla_\xi \tau)^2 +
\epsilon^{2p} |\boldsymbol{\nabla}_\perp \tau|^2} = 1.
\end{split}
\end{equation}
For $\tfrac{1}{2} < p < 1$, this result (multiplied by $\epsilon^{-2p}$)
simplifies considerably as $\epsilon \to 0$:
\begin{equation}
\label{kpz}
\begin{split}
\nabla_\xi \tau + \tfrac{1}{2} |\boldsymbol{\nabla}_\perp \tau|^2 =
{}&{-}\epsilon^{1-2p}\, U_\parallel(0,\epsilon^{-p} \xi,\mathbf{x}_\perp)\\
&- \epsilon^{1-2p}\, V(0,\epsilon^{-p} \xi,\mathbf{x}_\perp)\\
\equiv {}&\eta(\xi,\mathbf{x}_\perp).
\end{split}
\end{equation}

\begin{figure}
\begin{center}
\includegraphics[width=88.5mm]{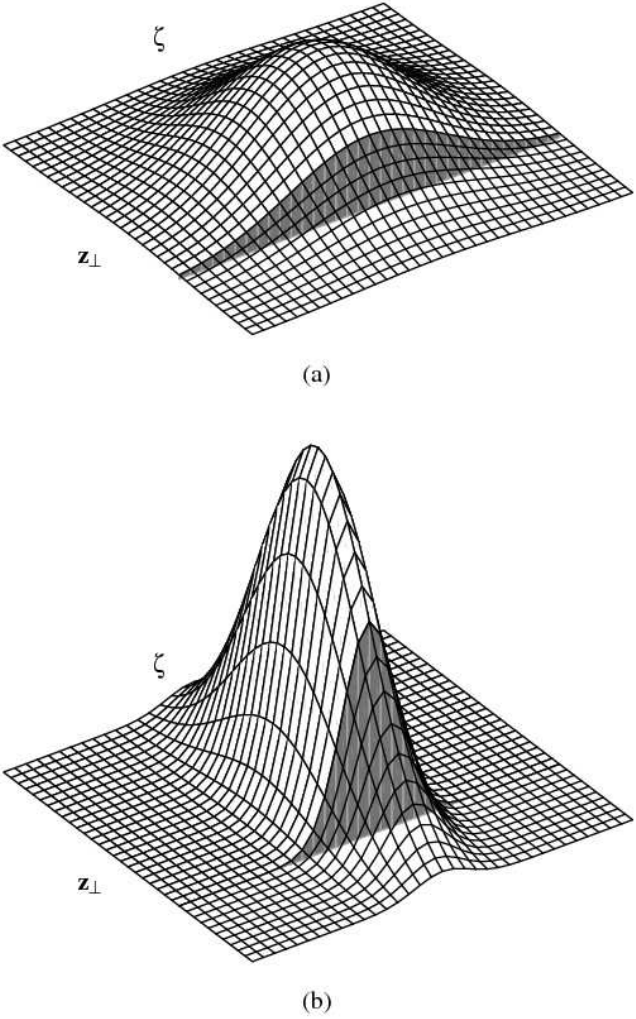}%
\end{center}
\caption{\label{white}Convergence of weak fluctuations to white noise. Choosing
a simple random medium with $\langle\gamma(\mathbf{0})\,
\gamma(\mathbf{z})\rangle \sim \exp(-z^2)$, we show the two-point correlation
function of $\eta$, Eq.\ (\protect\ref{gg}), for $\epsilon = 1$ (a) and
$\epsilon = 0.2$ (b). As $\epsilon \to 0$, the function is infinitely
compressed in the ``time'' direction $\zeta$, but with the choice of scaling
exponent $p = \tfrac{2}{3}$, its time integral at a given $\mathbf{z}_\perp$
(shaded area) remains fixed. This integral gives the
$\mathbf{z}_\perp$-dependent coefficient of a delta function that describes
spatially correlated white-in-time noise.}
\end{figure}

As expected, both the component $\mathbf{U}_\perp$ and the time dependence of
the medium go away in this limit. The front responds to the frozen random field
$\gamma(x_\parallel,\mathbf{x}_\perp) \equiv -(U_\parallel + V)$, which is
homogeneous but not generally isotropic because a specific component
of~$\mathbf{U}$ is selected. The rescaled function $\eta(\xi,\mathbf{x}_\perp)
= \epsilon^{1-2p}\, \gamma(\epsilon^{-p} \xi,\mathbf{x}_\perp)$ appears to
diverge as $\epsilon \to 0$, but its dependence on the ``time''~$\xi$ also
becomes more and more rapid, and in fact with $p = \tfrac{2}{3}$ it converges
under mild assumptions to a white-noise process in~$\xi$. The two-point
correlation function of $\eta$ is
\begin{equation}
\label{gg}
\begin{split}
\langle\eta(\xi,\mathbf{x}_\perp)\, \eta(\xi &+ \zeta,\mathbf{x}_\perp +
\mathbf{z}_\perp)\rangle = \langle\eta(0,\mathbf{0})\,
\eta(\zeta,\mathbf{z}_\perp)\rangle\\
= {}&\epsilon^{2-3p} \epsilon^{-p} \langle\gamma(0,\mathbf{0})\,
\gamma(\epsilon^{-p} \zeta,\mathbf{z}_\perp)\rangle\\
\xrightarrow{\epsilon \to 0} {}&\epsilon^{2-3p}\, \delta(\zeta)
\int_{-\infty}^\infty \D a\, \langle\gamma(0,\mathbf{0})\,
\gamma(a,\mathbf{z}_\perp)\rangle,
\end{split}
\end{equation}
where the rescaled expression is squeezed into a Dirac delta function in $\xi$,
corresponding to white noise, times a factor that remains finite for $p =
\tfrac{2}{3}$. Because any interval of~$\xi$ corresponds to an infinitely long
sample of the medium, $\eta$ is a Gaussian random field even if $\gamma$ is
not, provided the correlations of $\gamma$ decay fast enough with~$x_\parallel$
so that the central limit theorem is applicable. (For example, in homogeneous
turbulence with strong intermittency, the statistics of $\gamma$ are highly
non-Gaussian; but because the flow decorrelates over a few integral scales, the
long-term effect on the front is given as $\epsilon \to 0$ by Gaussian white
noise.) Despite becoming white in ``time'', the $\eta$ correlation function
(\ref{gg}) remains smooth in ``space''; this $\epsilon \to 0$ limit is
visualized in Fig.~\ref{white}.

To confirm that Eq.\ (\ref{kpz}) with $p = \tfrac{2}{3}$ describes a Burgers
fluid, note that
\begin{equation}
\boldsymbol{\nabla}T_0 = (1 + \epsilon^{4/3} \nabla_\xi \tau, \epsilon^{2/3}
\boldsymbol{\nabla}_\perp \tau),
\end{equation}
and so the physical particle velocity (\ref{dxdt1}) gives
\begin{equation}
\begin{split}
\mathbf{w} \equiv {}&\frac{\D\mathbf{x}_\perp}{\D\xi} = \epsilon^{-2/3}\,
\frac{\D\mathbf{x}_\perp/\D t}{\D x_\parallel/\D t}\\
= {}&\epsilon^{-2/3} \frac{\epsilon \mathbf{U}_\perp + (1 + \epsilon V)
\epsilon^{2/3} \boldsymbol{\nabla}_\perp
\tau/|\boldsymbol{\nabla}T_0|}{\epsilon U_\parallel + (1 + \epsilon V)(1 +
\epsilon^{4/3} \nabla_\xi \tau)/|\boldsymbol{\nabla}T_0|}\\
\xrightarrow{\epsilon \to 0} {}&\boldsymbol{\nabla}_\perp \tau
\end{split}
\end{equation}
for the Burgers ``velocity''~$\mathbf{w}$ in the time variable $\xi$. The
$\mathbf{x}_\perp$ gradient of Eq.\ (\ref{kpz}) then yields
\begin{equation}
\label{burg}
\nabla_\xi \mathbf{w} + (\mathbf{w} \cdot \boldsymbol{\nabla}_\perp) \mathbf{w}
= \boldsymbol{\nabla}_\perp \eta(\xi,\mathbf{x}_\perp),
\end{equation}
which is the inviscid Burgers equation with white-noise forcing \cite{IK03}.
Since the fluid elements are precisely the front-tracking particles, cusp
singularities in the front are described by the standard jump conditions for
Burgers shocks \cite{L73}.

Because $\eta$ averages to zero, Eq.\ (\ref{kpz}) implies that in a steady
state
\begin{equation}
C \equiv -\langle\nabla_\xi \tau\rangle = \tfrac{1}{2}
\langle|\boldsymbol{\nabla}_\perp \tau|^2\rangle = \tfrac{1}{2} \langle
w^2\rangle.
\end{equation}
Thus the prefactor of $\epsilon^{4/3}$ in the front speedup is the steady-state
energy density of the white-noise-driven Burgers fluid. It was shown rigorously
\cite{IK03} that for white noise that is periodic in space with any given
smooth correlation function, there exists a unique statistically steady state
for the inviscid Burgers equation. This holds in any spatial dimension,
generalizing a previous result \cite{EKMS00} for the one-dimensional Burgers
equation (corresponding to two-dimensional propagation). Since an assumption of
spatial periodicity is an accepted device for treating bulk properties, we have
convincing evidence that $C$~is well defined for a realistic smooth random
medium given its two-point spatial correlation function $f(\mathbf{z}) \equiv
\langle\gamma(\mathbf{0})\, \gamma(\mathbf{z})\rangle$, which determines the
$\eta$ correlation function (\ref{gg}), independent of higher statistics.

For geometrical optics in a quenched random medium, the essential rescaling
technique and a reduction to white noise were previously demonstrated at the
level of the ray equations \cite{W84,NW91}, but the first-passage problem and
the disappearance of rays at cusps (leading to the inviscid Burgers equation)
were not considered. On the other hand, the analogous shock singularities of
the Burgers fluid can be eliminated by a variant of Eq.\ (\ref{burg}) with an
additional term $-\nu\nabla_\perp^2 \mathbf{w}$ on the left, known as the
viscous Burgers equation \cite{BMP95}. (The viscosity $\nu > 0$ smooths the
shocks.) The corresponding variant of Eq.\ (\ref{kpz}), with
$-\nu\nabla_\perp^2 \tau$ on the left, is a form of the Kardar-Parisi-Zhang
(KPZ) equation for interface growth \cite{KPZ86}. Under forcing that is
continuous in space and time, the solution of these equations is known to
approach as $\nu \to 0$ a limiting ``viscosity solution'' that reproduces the
inviscid shocks \cite{CL83}. Under white-noise forcing, the steady state of the
viscous Burgers equation was analyzed using field theory \cite{P95,BMP95}, but
the $\nu \to 0$ limit is more subtle. Fortunately, a recent theorem
\cite{GIKP05} establishes that the viscous white-noise steady state converges
to the inviscid one, making the $\nu \to 0$ results of field theory applicable
to the $\epsilon \to 0$ limit of Huygens propagation. A previous description of
weakly advected flames based on the KPZ equation \cite{F95} assumed a
white-noise velocity field for convenience, but did not observe that it could
arise from a realistic smooth medium by rescaling, yielding the
$\epsilon^{4/3}$ law in the process.

\section{\label{Concl}Conclusion}

The newly established equivalence between $\nu \to 0$ Burgers turbulence and
the $\epsilon \to 0$ propagation problem allows adaptation of field-theoretic
\cite{BMP95} and numerical \cite{GK98} methods previously applied to the former
in order to analyze the latter. Results of this adaptation will be reported
elsewhere. The results obtained here already indicate the central importance of
the spatial structure of the random medium to front propagation. Indeed, a
sufficiently anisotropic medium can have a different scaling law, explaining
the $\epsilon^2$ dependence of the speedup derived for one such medium
\cite{AB03}. There, a flow $\mathbf{U}$ is constructed by adding (with random
phases) a finite number of Fourier modes, none of whose wavevectors are exactly
orthogonal to $x_\parallel$. In this homogeneous but anisotropic medium, the
integral in Eq.\ (\ref{gg}) oscillates and averages to zero. The amplitude of
the white noise then vanishes and so $C = 0$, indicating correctly that the
speedup goes to zero faster than $\epsilon^{4/3}$. As we will subsequently
report, $\epsilon^2$ scaling in this random medium is a direct result of the
gross anisotropy---similar to the original argument for $\epsilon^2$ scaling
\cite{CW79}, which was validated for an $x_\parallel$-periodic medium
\cite{AS91}. We have here established $\epsilon^{4/3}$ scaling as the generic
case, applying to random media with a finite spectral density orthogonal to
$x_\parallel$, such as isotropic media. (Although isotropic flows $\mathbf{U}$
lead to anisotropic $\gamma$, the generic scaling still holds because in
realistic cases the relevant transverse modes of $U_\parallel$ have a finite
spectral density; in particular, they are not constrained even if
incompressibility is assumed. Furthermore, generic quantitative deviations from
flow isotropy that maintain this finiteness do not alter the scaling.)

There is a different sense in which $\epsilon^2$ scaling exists even for
isotropic media: as a transient starting from a flat initial front. Before
cusps form, a linearized approximation is adequate; we expect front tilts
proportional to $\epsilon$ and thus a speedup proportional to $\epsilon^2$,
both systematically increasing with time. In the Burgers picture, there is an
order-unity interval of $\xi$ before shocks become important, during which the
steady energy input results in an energy density proportional to $\xi$. The
transient speedup is therefore proportional to $\epsilon^{4/3} \xi = \epsilon^2
x_\parallel$. As stated previously, our analysis applies after a distance
$x_\parallel \sim \epsilon^{-2/3}$, where a steady-state $\epsilon^{4/3}$
speedup is attained.

For premixed combustion applications, while weak advection is a useful limiting
case, the strong-advection limit $|\mathbf{u}| \gg v$ is more important
\cite{W85}. Because the white-noise reduction and effective freezing of the
medium no longer apply, we expect the strongly turbulent burning velocity to
depend not only on the two-point spatial correlation function $f$ but also on
more complicated spatiotemporal flow statistics. A widely used
isotropic-turbulence burning-velocity model \cite{Y88} fails to incorporate
this nonuniversality and also reduces to $\epsilon^2$ instead of
$\epsilon^{4/3}$ scaling in the weak limit.

A further implication of our analysis for strong turbulence comes from a
monotonicity property: For a given initial front and flow
$\mathbf{u}(t,\mathbf{x})$, all fluid elements burned by a later time for
laminar flame speed $v_1$ are also burned for $v_2 > v_1$. Thus the turbulent
burning velocity can only increase with $v$, even into the weak-advection
regime $v \gg |\mathbf{u}|$. The aforementioned field-theoretic constraints on
the weakly turbulent burning velocity then establish useful boundedness
properties of the strongly turbulent one. Finally, although we derived the
trajectory equations (\ref{dxdt}) and (\ref{dndt}) assuming $|\mathbf{u}| < v$,
they are Galilean invariant and are thus valid in general, because Huygens'
principle is local and we can switch to a reference frame where $|\mathbf{u}| <
v$ in a neighborhood of a given event. The trajectory equations do not require
a continuous representation of the front for constructing the
normal~$\mathbf{n}$ (except initially) and therefore may be useful in numerical
studies at all turbulence intensities, although for very low intensity it is
advantageous to exploit the Burgers equivalence.

\begin{ack}
The US Department of Energy, Office of Basic Energy Sciences, Division of
Chemical Sciences, Geosciences and Biosciences supported this work. Sandia is a
multiprogram laboratory operated by Sandia Corporation, a Lockheed Martin
Company, for the US Department of Energy under contract DE-AC04-94AL85000.
\end{ack}

\appendix

\section{\label{Law}Law of motion for first-passage trajectories}

The first arrival at an arbitrary point from an initial front is a constrained
optimum: the result of minimizing the travel time $T$ among virtual
trajectories $\mathbf{x}(t)$ obeying Eq.\ (\ref{huygens}), or equivalently
(\ref{dxdt}) with arbitrary unit $\mathbf{n}(t)$, such that $\mathbf{a} \equiv
\mathbf{x}(0)$ is on the initial front and $\mathbf{y} \equiv \mathbf{x}(T)$ is
the desired endpoint. Using the method of Lagrange multipliers, we eliminate
some of the constraints by introducing auxiliary variables. Consider a
trajectory $\mathbf{x}(t)$ obeying the constraints. If the travel-time
variation $\delta T$ vanishes for all infinitesimal trajectory variations
$\delta\mathbf{x}(t)$ that preserve the constraints (a necessary condition for
a minimum), then for trajectory variations that keep $\mathbf{a}$ on the
initial front but are otherwise arbitrary, $\delta T$ must be proportional to
the deviations from the other constraints:
\begin{equation}
\label{dT}
\delta T = \delta\mathbf{y} \cdot \boldsymbol{\kappa} + \int_0^T\D t\,
\delta[\mathbf{u}(t,\mathbf{x}) + v(t,\mathbf{x})\, \mathbf{n}(t) -
\dot{\mathbf{x}}(t)] \cdot \boldsymbol{\lambda}(t).
\end{equation}
Here the overdot denotes a time derivative, $\boldsymbol{\kappa}$~is the
Lagrange multiplier for the constraint on the endpoint $\mathbf{y}$ and
$\boldsymbol{\lambda}(t)$ is the Lagrange multiplier for the time-dependent
constraint (\ref{dxdt}). Equation (\ref{dT}) must hold for all infinitesimal
variations $\delta T$, $\delta\mathbf{x}(t)$, $\delta\mathbf{n}(t)$ as long as
the starting point $\mathbf{a}$ remains on the initial front and
$\mathbf{n}(t)$ remains a unit vector.

The total variation of $\mathbf{y} = \mathbf{x}(T)$ is $\delta\mathbf{y} =
\delta\mathbf{x}(T) + \dot{\mathbf{x}}(T)\, \delta T$ (i.e., the endpoint can
be changed either by altering the trajectory itself or by following it for a
longer or shorter time). Accordingly, Eq.\ (\ref{dT}) becomes
\begin{alignat}{2}
\delta T = {}&[\delta\mathbf{x}(T) + \dot{\mathbf{x}}(T)\, \delta T] \cdot
\boldsymbol{\kappa}\span\span\nonumber\\
&+ \int_0^T\D t\, [&&\delta\mathbf{x}(t) \cdot \LRVec A(t,\mathbf{x}) +
\delta\mathbf{x}(t) \cdot \boldsymbol{\nabla}v(t,\mathbf{x})\,
\mathbf{n}(t)\nonumber\\
&&&+ v(t,\mathbf{x})\, \delta\mathbf{n}(t) - \delta\dot{\mathbf{x}}(t)] \cdot
\boldsymbol{\lambda}(t).
\end{alignat}
Because $\delta\mathbf{n}$ is constrained only by $\delta\mathbf{n} \cdot
\mathbf{n} = 0$, the term proportional to $\delta\mathbf{n} \cdot
\boldsymbol{\lambda}$ shows that $\boldsymbol{\lambda}$ must be along
$\mathbf{n}$, say $\boldsymbol{\lambda}(t) = \mathbf{n}(t)\, \mu(t)$. Next we
integrate $-\delta\dot{\mathbf{x}} \cdot \boldsymbol{\lambda}$ by parts to
obtain
\begin{alignat}{2}
\delta T = {}&\dot{\mathbf{x}}(T) \cdot \boldsymbol{\kappa}\, \delta T +
\delta\mathbf{x}(T) \cdot [\boldsymbol{\kappa} - \mathbf{n}(T)\,
\mu(T)]\span\span\nonumber\\
&+ \delta\mathbf{a} \cdot \mathbf{n}(0)\, \mu(0)\span\span\nonumber\\
&+ \int_0^T\D t\, \delta\mathbf{x}(t) \cdot [&&\LRVec A(t,\mathbf{x}) \cdot
\mathbf{n}(t)\, \mu(t) + \boldsymbol{\nabla}v(t,\mathbf{x})\, \mu(t)\nonumber\\
\label{dT1}
&&&+ \dot{\mathbf{n}}(t)\, \mu(t) + \mathbf{n}(t)\, \dot\mu(t)].
\end{alignat}

Validity of Eq.\ (\ref{dT1}) for arbitrary $\delta T$ gives
\begin{equation}
\label{uvk}
1 = \dot{\mathbf{x}}(T) \cdot \boldsymbol{\kappa} = [\mathbf{u}(T,\mathbf{y}) +
v(T,\mathbf{y})\, \mathbf{n}(T)] \cdot \boldsymbol{\kappa}.
\end{equation}
Validity for arbitrary $\delta\mathbf{x}(T)$ gives
\begin{equation}
\label{knm}
\boldsymbol{\kappa} = \mathbf{n}(T)\, \mu(T).
\end{equation}
Validity for arbitrary $\delta\mathbf{a}$ tangent to the initial front shows
that $\mathbf{n}(0)$ is in the normal direction, as claimed. Because any
instant of time can be considered ``initial'', a first-passage trajectory
always has $\mathbf{n}$ normal to the evolving front (while the trajectory
remains on the front). Finally, validity for arbitrary $\delta\mathbf{x}(t)$
gives (upon projecting orthogonal to $\mathbf{n}$)
\begin{equation}
\dot{\mathbf{n}}(t) = -P_{\mathbf{n}(t)}\bigl(\LRVec A(t,\mathbf{x}) \cdot
\mathbf{n}(t) + \boldsymbol{\nabla}v(t,\mathbf{x})\bigr),
\end{equation}
which justifies Eq.\ (\ref{dndt}) and, together with Eq.\ (\ref{dxdt}), defines
the law of motion. Results equivalent to Eqs.\ (\ref{dxdt}) and (\ref{dndt})
were obtained in geometrical acoustics by assuming, rather than proving, that
$\mathbf{n}$ is the front normal \cite{E74}.

Our variational method is reminiscent of the reciprocal Hamilton principle
\cite{GKN96}, a modern contribution to the classical mechanics of particles.
Whereas the ordinary Hamilton least-action principle requires the action to be
stationary under variation of a trajectory between two points over a fixed time
interval, the reciprocal Hamilton principle requires the travel time to be
stationary under variation of a trajectory with fixed action. Both principles
are equivalent to Newton's or Hamilton's equations of motion. In our problem,
Eqs.\ (\ref{dxdt}) and (\ref{dndt}) can in fact be derived from a Hamiltonian
\begin{equation}
\label{H}
H(t,\mathbf{x},\boldsymbol{\lambda}) = \mathbf{u}(t,\mathbf{x}) \cdot
\boldsymbol{\lambda} + v(t,\mathbf{x})\, |\boldsymbol{\lambda}|,
\end{equation}
if we treat $\boldsymbol{\lambda}$ (previously a Lagrange multiplier) as the
canonical momentum and define $\mathbf{n} \equiv
\boldsymbol{\lambda}/|\boldsymbol{\lambda}|$. The corresponding action is
\begin{equation}
\label{S}
\begin{split}
S &\equiv \int_0^T \D t\, [\boldsymbol{\lambda}(t) \cdot \dot{\mathbf{x}}(t) -
H(t,\mathbf{x},\boldsymbol{\lambda})]\\
&= \int_0^T \D t\, [\dot{\mathbf{x}}(t) - \mathbf{u}(t,\mathbf{x}) -
v(t,\mathbf{x})\, \mathbf{n}(t)] \cdot \boldsymbol{\lambda}(t),
\end{split}
\end{equation}
which vanishes for ``Huygens'' trajectories that obey Eq.\ (\ref{dxdt}), as
well as for many other trajectories that do not.

The reciprocal Hamilton principle implies that a first-passage trajectory has
stationary travel time not only among Huygens trajectories but also among the
wider class with $S = 0$. One might suppose we could consider the trajectory
with absolute minimum (hence stationary) travel time in the $S = 0$ class,
apply the reciprocal Hamilton principle to obtain the equations of motion
(\ref{dxdt}) and (\ref{dndt}) and then observe that such a trajectory obeys the
fundamental constraint (\ref{huygens}) of Huygens propagation and thus
represents the physical first passage. But this shortcut fails because of an
important technicality: Among trajectories with $S = 0$, there is no absolute
minimum travel time between two points. For simplicity, take a quenched medium
with $\mathbf{u} \equiv \mathbf{0}$. A sufficient condition for $S = 0$ is that
$\dot{\mathbf{x}} \cdot \boldsymbol{\lambda} = v|\boldsymbol{\lambda}|$, i.e.,
the component of $\dot{\mathbf{x}}$ along $\boldsymbol{\lambda}$ equals $v$. A
suitable $\boldsymbol{\lambda}$ exists as long as $|\dot{\mathbf{x}}| \ge v$,
and so the $S = 0$ class includes arbitrarily fast trajectories. Therefore the
reciprocal Hamilton principle is not a sound basis for deriving the
first-passage law of motion, although that law can be expressed in Hamiltonian
form. Instead, we have derived the law of motion from the correctly constrained
variational principle.

\section{\label{Eikonal}Generalized eikonal equation}

The extra implications of our variational principle give a useful result for
the spatial dependence of the first-passage time $T_0(\mathbf{y})$. Equation
(\ref{dT}) shows that for variations restricted to first-passage trajectories
($T = T_0$), so that the variation inside the integral is zero, we have $\delta
T_0 = \boldsymbol{\kappa} \cdot \delta\mathbf{y}$ and thus
$\boldsymbol{\nabla}T_0 = \boldsymbol{\kappa}$. Equation (\ref{knm}) and
$|\mathbf{n}| = 1$ then give
\begin{equation}
\mathbf{n} = \frac{\boldsymbol{\nabla}T_0}{|\boldsymbol{\nabla}T_0|},
\end{equation}
confirming that $\mathbf{n}$ is always normal to the front. Substituting into
Eq.\ (\ref{uvk}), we find
\begin{equation}
\label{uvT}
\mathbf{u}(T_0,\mathbf{y}) \cdot \boldsymbol{\nabla}T_0 + v(T_0,\mathbf{y})\,
|\boldsymbol{\nabla}T_0| = 1.
\end{equation}
A form of this equation is known in geometrical acoustics \cite{P89} as a
generalization of the ``eikonal'' equation of geometrical optics. We see that
it follows directly from Huygens' principle in a time-dependent advected
medium.

In fact, a closely related ``$G$~equation'' was obtained for idealized premixed
combustion \cite{KAW88,W85a}:
\begin{equation}
\dot G + \mathbf{u} \cdot \boldsymbol{\nabla}G = v |\boldsymbol{\nabla}G|,
\end{equation}
where a particular ``level surface'' of the function~$G$, say the set of points
$\mathbf{y}$ where $G(t,\mathbf{y}) = 0$, represents the front at time~$t$. The
gradient of the identity $G\bigl(T_0(\mathbf{y}),\mathbf{y}\bigr) = 0$ gives
$\boldsymbol{\nabla}T_0 = -\boldsymbol{\nabla}G/\dot G$ and proves the
equivalence with Eq.\ (\ref{uvT}). The $G$~equation corresponds to the
Hamilton-Jacobi equation of classical mechanics \cite{GPS02}, with Hamiltonian
(\ref{H}), momentum $\boldsymbol{\lambda} = -\boldsymbol{\nabla}G$ and
Hamilton's principal function $-G$. Although the equations of motion for this
Hamiltonian reproduce (\ref{dxdt}) and (\ref{dndt}), in general such
trajectories (``characteristics'' of the Hamilton-Jacobi equation) do not move
with level surfaces of Hamilton's principal function. They do so for this
Hamiltonian because the action (\ref{S}) vanishes, reflecting the equality of
group and phase velocity (absence of dispersion) in Huygens propagation.

\end{document}